\documentclass[twocolumn,prb,showpacs]{revtex4}
\usepackage{graphicx}
\usepackage{dcolumn}
\usepackage{bm}

\begin{document}
\title{Elastic Light Scattering by Semiconductor Quantum Dots}
\author {I.G.Lang, L.I.Korovin}
\affiliation{A. F. Ioffe Physical-Technical Institute, Russian
Academy of Sciences, 194021 St. Petersburg, Russia}
\author{S.T.Pavlov\dag\ddag}
\address{\dag Facultad de Fisica de la UAZ, Apartado Postal C-580, 98060 Zacatecas, Zac., Mexico \\
\ddag P. N. Lebedev Physical Institute, Russian Academy of
Sciences, 119991 Moscow, Russia}

\begin{abstract}
Elastic light scattering by low-dimensional semiconductor objects
is investigated theoretically. The differential cross section of
resonant light scattering on excitons in quantum dots is
calculated.  The polarization and angular distribution of
scattered light do not depend on the quantum-dot form, sizes and
potential configuration if light wave lengths exceed considerably
the quantum-dot size. In this case the magnitude of the total
light scattering cross section  does not depend on quantum-dot
sizes. The resonant total light scattering cross section is about
a square of light wave length if the exciton radiative broadening
exceeds the nonradiative broadening. Radiative broadenings are
calculated.
\end{abstract}


\maketitle

\section{Introduction}

Measurement of elastic light scattering by size-quantized
low-dimensional semiconductor objects (quantum wells, quantum
wires and quantum dots) is a simple and convenient method to
investigate excitonic excitations in these objects.

If exciton energy levels are discrete, light scattering becomes
stronger resonantly at coincidence of the light frequency
 $ \omega_l $ with the exciton frequency $
\omega_0 $. A resonant peak width is determined by the exciton
damping $ \Gamma $. The same concerns to the light absorption.

For the first time the role of so-called radiative damping $
\gamma_r $ of excitons  was discovered in light reflection of
quantum wells. It was shown that the total damping $ \Gamma $
consists of two parts: $ \Gamma =\gamma_r +\gamma, $ where $
\gamma $ is the exciton nonradiative damping [1].  The same
concept was extended on light absorption by quantum wells [2] (see
also review [3]). The light reflection by quantum wells, quantum
wires and quantum dots was considered for the first time in [4].

 There are two methods to investigate theoretically the light scattering by semiconductor
objects. First of them we name as semiclassical method, since it
is reduced to calculation of classic electric fields, whereas the
description of electronic systems is quantum (it needs to be
mentioned that a creation of an electron-hole pair is described by
a non-diagonal matrix elements ${\bf p} _ {cv} $ of the momentum
operator). The semiclassical method is described in [5]. It
consists of calculation of averaged (on the ground state of a
crystal) current and charge densities induced by a stimulating
electric field and solution of the Maxwell equations inside and
outside of the object with subsequent use of boundary conditions
for electric and magnetic fields. Inelastic light scattering (for
example, the Raman scattering) is caused by fluctuations of
currents and charges. Let us emphasize that calculations of
current and charge densities are made in view of exciton
nonradiative broadenings $ \gamma $ [5], that allows to calculate
not only scattering, but the light absorption also (at $ \gamma=0
$ the light absorption is absent). The quasi-classical method in
calculations of reflection and absorption coefficients in quantum
wells was used in [4,6-8]. In [9] the same method was applied to
calculation of electric fields arising at resonant light
scattering on excitons $ \Gamma_6\times\Gamma_7 $ in spherical
quantum dots, consisting of crystals of a class $T_d $ (for
example, GaAs) and limited by an infinitely high potential
barrier.

The second method is quantum one. The electric field
 is quantized and the quantum perturbation theory is used. We have checked
up that both methods give identical results for dimensionless
light reflection coefficients in quantum wells if the
light-exciton interaction is taken into account in the lowest
order. That is admitted at $ \gamma_r\ll \gamma $.

Certainly, the quasi-classical method has a number of advantages
before the quantum method. First, it allows to take into account
precisely the light-electron interaction, i.e., all processes of
light reradiation and reabsorption. The exact description is being
achieved, if to substitute genuine electric fields in expressions
for current and charge densities. Then we obtain automatically
that the resonant contributions of excitons in dimensionless light
reflection and absorption coefficients in quantum wells and in the
light scattering cross sections in quantum wires and quantum dots
contain factors
 $[(\omega_l-\tilde{\omega}_0)^2+\Gamma^2/4]^{-1}$, where
$ \tilde {\omega} _0 =\omega_0 +\Delta\omega $ is the renormalized
exciton frequency [6,8,9].

Secondly, the quasi-classical method allows to calculate precisely
the light absorption by low-dimensional objects. For quantum wells
this task is solved in [6-8]. In [9] $ \gamma=0 $, and the light
absorption by a quantum dot is absent.

Thirdly, the quasi-classical method allows to go from a
monochromatic irradiation to a pulse irradiation and to obtain new
interesting results [10-14].

Finally, the quasi-classical method allows to take into account a
permittivity difference of objects and environment [6,9,12].

However, the quantum method has one decisive advantage. - It is
much easier especially for quantum wires and QDs, that is shown
below.

\section{The quantum theory}
Let us calculate a probability of a stimulating photon absorption
and a scattered photon creation. According to the perturbation
theory
\begin {equation}
\label {l} W_l={2\pi\over\hbar}\sum_f|M_{fi}|^2\delta(E_f-E_i),
\end {equation}
where $E_i (E_f) $ is the initial (final) state energy,
\begin {equation}
\label {2} M _ {f i} = \sum_m {\langle f|V|m\rangle\langle
m|V|i\rangle\over E_i-E_m+i\delta\hbar}
\end {equation}
is the compound matrix element, $E_m $ is the intermediate state
energy, $ \delta\rightarrow +0 $.

The charge - electric field interaction is as follows
\begin {equation}
\label {3} V =-\int d^3r\, {\bf d} ({\bf r}) {\bf E} ({\bf r}),
\end {equation}
where the polarization density
\begin {equation}
\label {4} {\bf d} ({\bf r}) = \sum_i {\bf r} _i\rho_i ({\bf r}),
~~\rho_i ({\bf r}) =e\delta ({\bf r} - {\bf r} _i)
\end {equation}
is introduced. We use the field $ {\bf E} ({\bf r}) $  in the
secondary quantization representation [15, p. 579]
\begin {eqnarray}
\label {5} &&{\bf E} ({\bf r}) ={i\over\nu}\sqrt{{2\pi\hbar\over
V_0}}\nonumber\\
&\times& \sum _ {{\bf k}, \mu}\sqrt{ \omega_k } (c _ {\mu {\bf k}}
{\bf e} _ {\mu {\bf k}} e ^ {i {{\bf k} {\bf r}}} -c ^ + _ {\mu
{\bf k}} {\bf e} ^ * _ {\mu {\bf k}} e ^ {-i {{\bf k} {\bf
r}}}),~~~
\end {eqnarray}
where $V_0 $ is the normalization volume, $ \omega=ck/\nu $ is the
frequency, $ {\bf k} $ is the light wave vector, $ \nu =\sqrt
{\varepsilon} $ is the light refraction coefficient, which is
 identical inside and outside of the semiconductor volume, $ {\bf
e} _ {\mu {\bf k}} $ is the polarization vector, $ \mu $ is the
polarization index, $c ^ + _ {\mu {\bf k}} (c _ {\mu {\bf k}}) $
is the photon creation (annihilation) operator. In (5), the
approximation $u _ {\mu {\bf k}} =c/\nu $ is used, where $u _ {\mu
{\bf k}} $ is the light group velocity. The field (5) is
normalized so that the energy in the volume $V_0 $ equals $ \sum _
{{\bf k}, \mu} \hbar\omega_k (N _ {\mu {\bf k}} +1/2) $.

In the effective mass approximation [5],
\begin {eqnarray}
\label {6} &&{\bf d} ({\bf r}) = {\bf d} ^ {nd} ({\bf r})\nonumber\\
&&= \sum_\eta [{\bf d} _ {cv\eta} F ^ *_\eta ({\bf r}) a ^ +_\eta
+ {\bf d} _ {cv\eta} ^ *F_\eta ({\bf r}) a_\eta],~~~
\end {eqnarray}
where superscript $nd $ means a non-diagonal part of the operator
(having only non-diagonal matrix elements), $a ^ +_\eta (a_\eta) $
is the exciton creation (annihilation) operator with an index $
\eta, F_\eta ({\bf r}) $ is the exciton envelope wave function at
$ {\bf r} _e = {\bf r} _h = {\bf r}, {\bf r} _e ({\bf r} _h) $ is
the electron (hole) radius - vector, $ {\bf d} _ {cv\eta} = -
(ie/m_0\omega_g) {\bf p} _ {cv\eta}$, $m_0 $ is the free electron
mass, $ \hbar\omega_g $ is the energy gap, $ {\bf p} _ {cv\eta} $
is the momentum interband matrix element. Using (5) and (6) in
(3), we obtain $ V=V_1+V_2+h.c.$, where
\begin {eqnarray}
 \label{7}V_1&=&-{e\over m_0\omega_g\nu} \left ({2\pi\hbar} \over
V_0\right) ^ {1/2} \sum_\eta\sum _ {{\bf k}, \mu} a ^ +_\eta
c _ {\mu {\bf k}} \omega_k ^ {1/2} \nonumber \\
&\times&({\bf p} _ {cv\eta} {\bf
e} _ {\mu {\bf k}}) P_\eta ^ * ({\bf k}), \nonumber \\
 V_2&=&{e\over m_0\omega_g\nu} \left ({2\pi\hbar} \over
V_0\right) ^ {1/2} \sum_\eta\sum _ {{\bf k}, \mu} a ^ +_\eta
c ^ + _ {\mu {\bf k}} \omega_k ^ {1/2} \nonumber \\
 &\times&({\bf p} _ {cv\eta} {\bf e} ^ * _ {\mu {\bf
k}}) P_\eta ^ * (- {\bf k}),
\end {eqnarray}
\begin {equation}
\label {8} P_\eta ({\bf k}) = \int d^3\, r e ^ {-i {\bf k} {\bf
r}} F_\eta ({\bf r}).
\end {equation}
The compound matrix element (2) consists of two parts $ M _ {fi}
=M _ {fi} ^1+M _ {fi} ^2$,
\begin {equation}
\label {9} M _ {f i} ^ {1 (2)} = \sum_m {\langle f|V ^ + _ {1 (2)}
|m\rangle\langle m|V _ {1 (2)} |i\rangle\over
E_i-E_m+i\delta\hbar}.
\end {equation}
In the initial state $ |i\rangle $, there are the semiconductor
ground state and $N_l $ photons with a wave vector $ {\bf k} _l $
and polarization $ {\bf e} _l $,
  $N_l\gg 1 $. In the final state $ |f\rangle $, there are
$N_l-1 $ photons of exciting light and one photon of scattered
light with a wave vector $ {\bf k} _s $ and polarization $ {\bf e}
_s $.

In an intermediate state (process 1), there are $ N_l-1 $ photons
of exciting light and an exciton $ \eta $; in the intermediate
state (process 2), there are $N_l $ photons of exciting light, one
photon of scattered light and an exciton $ \eta $.

For probability $W_l $ we obtain the result
\begin {eqnarray}
 \label{10}W_l={(2\pi)^3\over V_0^2\hbar^2} \left ({e^2\over
 m_0^2\omega_g^2}\right)^2{N_l
 \omega_l\omega_s\over\nu^4}\nonumber\\
\times\sum _ {{\bf k} _s, \mu} \left |\sum_\eta {\tilde A} _
\eta\right | ^ 2 \delta (\omega_l-\omega_s),
\end {eqnarray}
where the designation
\begin {eqnarray}
\label {11} {\tilde A} _ \eta = {({\bf p} _ {cv\eta} {\bf e} _l)
({\bf p} _ {cv\eta} {\bf e} _s) ^ *P_\eta ^ * ({\bf k} _l) P_\eta
({\bf k} _s) \over
\omega_l-\omega_\eta+i\delta} \nonumber \\
- {({\bf p} _ {cv\eta} ^ * {\bf e} _l) ({\bf p} _ {cv\eta} ^ *
{\bf e} _s) ^ *P_\eta (- {\bf k} _l) P ^ *_\eta (- {\bf k} _s)
\over \omega_l +\omega_\eta+i\delta}~~~
\end {eqnarray}
is introduced. The summation in the RHS of (10) is carried out on
wave vectors $ {\bf k} _s $ and polarizations $ \mu $ of scattered
light.

Let us consider the resonant scattering, when energies $
\hbar\omega_l $ and $ \hbar\omega_\eta $ exceed slightly the
energy gap. Then the "non-resonant" second term in the RHS of (11)
must be omitted.

Substituting the summation $ {\bf k} _s $ by the integration and
integration on the module $ k_s $ by the integration on frequency
$ \omega_s=ck_s/\nu $, we obtain
\begin {eqnarray}
\label {12} W_l =\left ({e^2\over m_0^2\omega_g^2} \right) ^2 {
N_l\omega_l^4\over V_0\hbar^2c^3\nu}\nonumber\\
 \times\sum_\mu\int do_s\left |\sum_\eta A_\eta\right | ^ 2,
\end {eqnarray}
where $A_\eta$ is the resonant term from (11). The expression (12)
is universal one: It is applicable to any low-dimensional
semiconductor objects, including the presence of quantizing
magnetic field.

\section{Light scattering on quantum dots.}
 A quantum dot may be of any
form (sphere, cube, disk) and be limited by any energy potential
 (parabolic, rectangular) of any height. All the features of
 quantum-dot structure influence only on $P_\eta ({\bf k}) $ for an exciton $
\eta $. The angular distribution of scattered light depends also
on the structure of vectors $ {\bf p} _ {cv\eta} $, which,
generally speaking, are complex. For cubic crystals (class $T_d $)
these  vectors are different for excitons, containing heavy or
light holes from the valence band, chipped off by the spin-orbital
interaction [16,17]. Under an exciton we consider any state of an
electron-hole pair with a discrete energy level.

In a case of a quantum dot, it is naturally to introduce the light
scattering cross section. According to Eq. (12), the scattered
energy flux in an solid angle interval $do_s $ in time unit equals
\begin {eqnarray}
\label {13} \hbar\omega_l dW_l =\left ({e^2\over m_0^2\omega_g^2}
\right) ^2 {N_l\omega_l^5\over V_0\hbar c^3\nu}\nonumber\\
\times\sum_\mu\left |\sum_\eta A_\eta\right | ^ 2do_s.
\end {eqnarray}

The  energy flux of stimulating light on the unit area in time
unit equals
\begin {equation}
\label {14} S_l = {N_l\hbar\omega_l\over V_0} {c\over\nu}.
\end {equation}
Eq. (13) divided by (14) is the light scattering differential
cross section
\begin {equation}
\label {15} d\sigma =\left ({e^2\over\hbar c} \right) ^2 {
\omega_l^4\over c^2\omega_g^4m_0^4}\sum_\mu\left|\sum_\eta
A_\eta\right | ^ 2do_s.
\end {equation}
Eq. (15) describes an angular dependence and (without sum on $ \mu
$) polarization of scattered radiation. If $ \omega_l $ is close
to $ \omega_\eta $ of some excitonic state, the resonant
amplification of scattered light is observed.

If the excitonic state (see section VI below) is degenerated
(i.e., the same energy $\omega_\eta=\omega_0$ corresponds to some
set of indexes $\eta$, and the function $P_\eta({\bf k})=P({\bf
k})$ and only  vectors ${\bf p}_{cv\eta}$ depend on $\eta$) the
contribution of this state into the light scattering section
equals
\begin {equation}
\label {16} {d\sigma_0\over do_s} = \sum_\mu {d\sigma _ \mu \over
do_s},
\end {equation}
\begin {eqnarray}
\label {17} {d\sigma_\mu\over do_s} &=& \left ({e^2\over\hbar c}
\right) ^2 {\omega_l^4\over c^2\omega_g^4m_0^4} \left |\Xi_\mu
({\bf e} _l {, \bf e} _s) \right | ^ 2\nonumber\\
&\times& {|P ({\bf k} _l) | ^2|P ({\bf k} _s) | ^2\over
(\omega_l-\omega_0) ^2 +\delta^2},
\end {eqnarray}
where
\begin {equation}
\label {18} \Xi_\mu ({\bf e} _l, {\bf e} _s) = \sum_\eta ({\bf p}
_ {cv\eta} {\bf e} _l) ({\bf p} _ {cv\eta} {\bf e} _s)^*.
\end {equation}
Let us designate by $R $ the greatest quantum-dot size and
consider a case, when a light wave length is much greater than $R
$, i.e.,  $k R\ll 1$. Then the value $ P ({\bf k} _l) \simeq P (0)
= \int d^3r F_\eta ({\bf r})$ does not depend on a wave vector $
{\bf k} $, and the resonant contribution to the cross section is
described by the formula (17), in which it is necessary to make
the replacement $|P ({\bf k} _l) | ^2|P ({\bf k} _s) | ^2 \simeq
|P (0) | ^4$.

It is possible to make the following conclusions. At $ k R\ll 1 $
the polarization and angular distribution of scattered light are
determined only by the factor $ | \Xi_\mu ({\bf e} _l, {\bf e} _s)
| ^2 $, containing the vector $ {\bf p} _ {cv\eta} $, i.e., it
does not depend on the quantum-dot form and on the exciton wave
function. The section magnitude does not depend on the quantum-dot
sizes. Certainly, the energy level $E_\eta $ position depends on
the quantum-dot form and sizes.

\section{The radiative broadening of the exciton level}
 It is well known (see, for example, [1-8]) that the exact
account of the  electron-electromagnetic field interaction and
account of the exciton nonradiative broadening $ \gamma_\eta $
 results in the replacement of the factor $(\omega_l-\omega_\eta+i\delta )^{-1}$
 by $[\omega_l- {\tilde\omega}_\eta+i(\gamma_{r\eta}+\gamma_\eta)/2]^{-1}$ in (11),
 where $ \gamma _ {r\eta} $ is the radiative broadening, $ \hbar {\tilde
\omega} _ \eta $ is the renormalized exciton energy, $ {\tilde
\omega}_\eta=\omega_\eta+\Delta\omega_\eta$.

The calculation of radiative broadening is made according to Eq.
(1). The matrix element $ M _ {f i} = \langle f|V|i\rangle$
corresponds to a direct transition from an initial state (in witch
there is the exciton $\eta$ , but the photons are absent) in final
state, containing the ground state of a crystal and a photon with
a wave vector $ {\bf k} $ and polarization $ \mu .$ Using Eqs. (3)
- (6), we obtain
\begin {eqnarray}
\label {19} \gamma_{r\eta}&=&{4\pi^2\over\hbar}{e^2\over
m_0^2\omega_g^2\nu^2V_0}\sum _ {{\bf k}, \mu} \omega_k | {\bf p} _
{cv\eta} {\bf e} _ {\mu {\bf k}} | ^2 \nonumber \\
&&\times |P_\eta ({\bf
 k})|^2\delta(\omega_\eta-\omega_k).
\end {eqnarray}
Having replaced summation on $ {\bf k} $ by integration, we obtain
the result
\begin {eqnarray}
\label {20}&& \gamma _ {r\eta} = {e^2\omega_\eta^3\nu\over
2\pi\hbar
m_0^2\omega_g^2c^3}\nonumber\\
&&\times\sum_\mu\int do _ {{\bf k} _ \eta} | {\bf p} _ {cv\eta}
{\bf e} _ {\mu {\bf k}_\eta }| ^2|P_\eta ({\bf k} _ \eta) |
^2,~~~~
\end {eqnarray}
where $ {\bf k_\eta} $ is the  vector with module $k_\eta =
\omega_\eta\nu/ c$. The formula (20) is applicable to any quantum
wells, quantum wires and quantum dots at any magnitudes of the
parameter $ k_\eta d $, where $d $ is the characteristic size.

 Under condition $ k_\eta R\ll 1 $, we obtain for a quantum dot
\begin {equation}
\label {21} \gamma_{r\eta}={e^2\omega_\eta^3\nu|P_\eta(0)|^2\over
2\pi\hbar m_0^2\omega_g^2c^3} \sum_\mu\int do _ {{\bf k} _ \eta} |
{\bf p} _ {cv\eta} {\bf e} _ {\mu {\bf k}_\eta} | ^2,
\end {equation}
whence it follows that the radiative broadening does not depend on
quantum-dot sizes.

\section{ Estimation of the magnitude of the light scattering section  in
the resonance} Taking into account the exciton energy corrections
and both radiative and nonradiative broadenings, with the help
(16) and (17) we obtain the formula for the total light scattering
cross section for any quantum dot near the resonance $ \omega_l =
{\tilde \omega} _0 $:
\begin {eqnarray}
 \label{22}&&\sigma_0=\left({e^2\over\hbar
 c}\right)^2{\omega_l^4\over\omega_g^4m_0^4c^2}\nonumber\\
 &\times&{|P({\bf
k} _l) | ^2 \over (\omega_l- {\tilde
 \omega}_0)^2+(\gamma_r+\gamma)^2/4}\nonumber\\
 &\times&\int do_s|P ({\bf k} _s) | ^2\sum_\mu | \Xi_\mu ({\bf
e} _l, {\bf e} _s) | ^2.
\end {eqnarray}
For an estimation of the magnitude of the radiative broadening, we
use Eq. (20). Let us assume that $ \gamma\ll\gamma_r $. Then, we
obtain
\begin {equation}
\label {23} \sigma_0 (\omega_l = {\tilde
 \omega}_0)={c^2\over\omega_l^2\nu^2}x={x\over\kappa_l^{2}},
\end {equation}
where $x $ is a number about unit.

It follows from (23) that the elastic light scattering cross
section in the resonance is about a square
 of wave length of the exciting light in the case, when the radiative
broadening exceeds nonradiative one. It can be carried out in
perfect quantum dots. This result is true  at $ k_lR\ll 1 $, and
at $ k_lR\geq 1 $ for any quantum dot, in which exciton energy
levels exist. Otherwise at $ \gamma\gg\gamma_r $ the magnitude of
the cross section in resonance decreases in $ (\gamma/\gamma_r) ^2
$ times.

\section{Excitons $ \Gamma_6\times\Gamma_7 $ in cubic crystals of the class $T_d $.}
As an example, we consider an exciton formed by an electron from
the twice degenerated conduction band $ \Gamma_6 $ and
 by a hole from the twice degenerated valence band $ \Gamma_7 $, chipped off by the spin-orbital interaction.
 Such exciton is considered in [9].

In designations of [17], electron wave functions have the
structure
\begin {equation}
\label {24} \Psi _ {e1} =i S\uparrow, ~~~~ \Psi _ {e1} =i
S\downarrow,
\end {equation}
and hole wave functions are
\begin {eqnarray}
 \label{25}\Psi_{h1}={1\over\sqrt{3}}(X-iY)\uparrow
- {1\over\sqrt {3}} Z\downarrow, \nonumber \\
 \Psi_{h2}={1\over\sqrt{3}}(X+iY)\downarrow
+ {1\over\sqrt {3}} Z\uparrow.
\end {eqnarray}
Combining functions (24) and (25) in pairs, we obtain the four
times degenerated excitonic states, for which the vector $ {\bf p}
_ {cv} $ equals
\begin {eqnarray}
\label {26} {\bf p}_{cv1}&=&{p_{cv}\over\sqrt{3}}({\bf e} _x -i
{\bf e} _y), \nonumber\\ { \bf
p}_{cv2}&=&{p_{cv}\over\sqrt{3}}({\bf e} _x +i {\bf
e} _y), \nonumber \\
{ \bf p}_{cv3}&=&{p_{cv}\over\sqrt{3}}{\bf e} _z, \nonumber\\ {
\bf p}_{cv4}&=&-{p_{cv}\over\sqrt{3}}{\bf e} _z,
\end {eqnarray}
where we have introduced a scalar $ p _ {cv} =i\langle S | {\hat
p} _x|X\rangle$. $ {\bf e} _x, {\bf e} _y, {\bf e} _z $ are unite
vectors along crystallographic axes.

Let us consider a circular polarization of exciting and scattered
light
\begin {eqnarray}
\label {27} {\bf e}_{l}^\pm&=&{1\over\sqrt{2}}({\bf e} _ {xl} \pm
i {\bf e} _ {yl}), \nonumber\\{ \bf e}_{s}^\pm
&=&{1\over\sqrt{2}}({\bf e} _ {xs} \pm i {\bf e} _ {ys}),
\end {eqnarray}
where unite vectors $ {\bf e} _ {xl} $ and $ {\bf e} _ {yl} $ are
perpendicular to the axis $z_l $ directed along the vector $ {\bf
k} _l $. Unite vectors $ {\bf e} _ {xs} $ and $ {\bf e} _ {ys} $
are perpendicular to the axis $z_s $ directed along the vector $
{\bf k} _s $.

Direction of $ {\bf k} _l $ concerning crystallographic axes is
arbitrarily. It is described by angles $ \vartheta_l, \varphi_l $,
where $ \vartheta_l $ is the angle between chosen crystallographic
axis $z $ and $ {\bf k} _l $. The direction of $ {\bf k} _s $ is
described by angles $ \vartheta_s, \varphi_s $. Direct calculation
of $\Xi ({\bf e} _ {l}, {\bf e} _ {s})$ (18) results in
\begin {eqnarray}
\label {28} &&\Xi({\bf e} _ {l} ^ +, {\bf e} _ {s} ^ +) = \Xi ^ *
({\bf e} _ {l} ^ -, { \bf e} _ {s} ^ -)\nonumber\\&=& {p _ {cv}
^2\over
 3}\{(1+\cos\vartheta_l\cos\vartheta_s)\cos(\varphi_s-\varphi_l)\nonumber\\
&+&\sin\vartheta_l\sin\vartheta_s\nonumber\\ &+&
 i(\cos\vartheta_l-\cos\vartheta_s)\sin(\varphi_s-\varphi_l)\},~~~
\end {eqnarray}
\begin {eqnarray}
\label {29} &&\Xi({\bf e} _ {l} ^ +, {\bf e} _ {s} ^ -) = \Xi ^ *
({\bf e} _ {l} ^ -, { \bf e} _ {s} ^ +) = \nonumber \\ &=& {p _
{cv} ^2\over
 3}\{(1-\cos\vartheta_l\cos\vartheta_s)\cos(\varphi_s-\varphi_l)\nonumber\\&-&\sin\vartheta_l\sin\vartheta_s
 \nonumber\\&+&  i(\cos\vartheta_l-\cos\vartheta_s)\sin(\varphi_s-\varphi_l)\}.~~~
\end {eqnarray}
Squaring (28) and (29) on the module, we obtain
\begin {eqnarray}
\label {30} | \Xi ({\bf e} _ {l} ^ +, {\bf e} _ {s} ^ +) | ^2 =
|\Xi ({\bf e} _ {l} ^ -, {\bf e} _ {s} ^ -) | ^2\nonumber\\ = {p _
{cv} ^4\over 9} (1 +\cos\theta)^ 2,
\end {eqnarray}
\begin {eqnarray}
\label {31} ||\Xi ({\bf e} _ {l} ^ +, {\bf e} _ {s} ^ -) | ^2 =
|\Xi ({\bf e} _ {l} ^ -, {\bf e} _ {s} ^ +) | ^2 |\nonumber\\ = {p
_ {cv} ^4\over 9} (1-\cos\theta)^ 2,
\end {eqnarray}
where $ \theta $ is the angle between $ {\bf k} _l $ and $ {\bf k}
_s $.

Let us notice that at a direction of light along the
crystallographic axis $z $, i.e.,
 $ {\bf k} _l $ along $z $, at polarization
$ {\bf e} _l ^ + $  only the exciton with $ \eta=1 $ and $ {\bf p}
_ {cv1} = (p _ {cv} /\sqrt {3}) ({\bf e} _x-i {\bf e} _y) $ is
excited, and at polarization $ {\bf e} _l ^- $  only exciton with
$ \eta=2 $ and $ {\bf p} _ {cv2} = (p _ {cv} /\sqrt {3}) ({\bf e}
_x+i {\bf e} _y) $ is excited. However, at any direction of light
concerning crystallographic axes, all four excitons are excited.

Substituting (30) and (31) in (17), we obtain  the differential
light scattering cross sections
\begin {equation}
\label {32} {d\sigma ^ {++} \over do_s} = {d\sigma ^ {--} \over
do_s} = \Sigma_0 {(1 +\cos\theta) ^2\over 9},
\end {equation}
\begin {equation}
\label {33} {d\sigma ^ {+-}\over do_s} = {d\sigma ^ {- +}\over
do_s} = \Sigma_0 {(1-\cos\theta) ^2\over 9},
\end {equation}
where superscript $ ++ $ means polarization of exciting
(scattered) light $ {\bf e} _l ^ + ({\bf e} _s ^+) $ etc.,
\begin {equation}
\label {34} \Sigma_0 =\left ({e^2\over\hbar c} \right)^2
{\omega_l^4 p _ {cv} ^4\over\omega_g^4m_0^4c^2} {|P ({\bf k} _l) |
^2|P ({\bf k} _s) | ^2\over ((\omega_l-\omega_0) ^2 +\delta^2)}.
\end {equation}
Summarizing on polarizations of scattered light, we obtain
\begin {equation}
\label {35} {d\sigma ^ {+}\over do_s} = {d\sigma ^ {-}\over do_s}
= {2\over 9}\Sigma_0 (1 +\cos^2\theta),
\end {equation}
where superscript $ + (-) $ means polarization of exciting light $
{\bf e} _l ^ + ({\bf e} _l ^-) $.

Finally, the total light scattering cross section $ \sigma ^ +
=\sigma ^- $ turns out as a result of integration on angles
determining a direction of
 $ {\bf k} _s $. The factor $ |P ({\bf k} _l) | ^2 $ can cause dependence of the cross section
 on a direction of exciting light, for example, if a quantum dot is a disk.

If the magnitude $P ({\bf k}) $ depends only on the module $ k $
\begin{equation}
\label{36} P ({\bf k} _l) =P ({\bf k} _s) = {\cal P} (k_l R),
\end{equation}
the magnitude of $ \Sigma_0 $ does not depend on a direction of
${\bf k}_l $ and $ {\bf k} _s $. But in this case (as it follows
from (34) and (35)) the light scattering is not isotropic. Under
condition (36), integration on angles, determining a direction of
$ {\bf k} _s $, is carried out easily, and we obtain the result
for the total section
\begin{eqnarray}
\label{37}\sigma^+=\sigma^-={32\pi\over 27}\left({e^2\over\hbar
c}\right)^2{\omega_l^4p_{cv}^4\over\omega_g^4m_0^4c^2}\nonumber\\
\times{|{\cal P}(k_lR)|^4\over(\omega_l-\omega_0)^2+\delta^2 }.
\end{eqnarray}
For example, using the envelope wave function
\begin {equation}
\label {38} F ({\bf r}) = {1\over 2\pi R} {\sin^2 (\pi r/R) \over
r^2} \theta (R-r),
\end {equation}
we obtain
\begin {eqnarray}
\label {39} {\cal P} (\kappa R) &=& {2\over \kappa R} \int_0 ^\pi
dx ~
 \sin\left ({\kappa Rx\over\pi} \right) {\sin^2x\over x}, \nonumber\\ {\cal P} (0) &=&1.
\end {eqnarray}
Under condition $ k_lR\ll 1 $ in Eqs. (34) and (37), we believe $P
({\bf k}) \simeq P (0) $ and obtain results applied in case of
small quantum dots of any form, sizes and potential configurations
for light scattering in cubic crystals of a class $T_d $, having
holes from the chipped off valence band. In cases of heavy or
light holes in the exciton structure, we obtain other results
caused by other structure of vectors $ {\bf p} _ {cv\eta} $.

At $ k_lR\geq 1$ it may be essentially to take into account a
difference of permittivities inside and outside of the quantum
dot.

Let us calculate the radiative broadening for excitons with
vectors $ {\bf p} _ {cv\eta} $ from (26).  The factors $ S _
{\eta\mu} = | {\bf p} _ {cv\eta} {\bf e} _ {\mu {\bf k}} | ^2$
enter in Eq. (21). Using (26), we obtain the following results
\begin {eqnarray}
\label {40} S _ {1 +} &=& S _ {2-} = {p _ {cv} ^2\over
6} (1 +\cos\vartheta) ^2, \nonumber \\
S _ {1-} &=& S _ {2 +} = {p _ {cv} ^2\over
6} (1-\cos\vartheta) ^2, \nonumber \\
 S_{3+}&=&S_{3-}=S_{4+}=S_{4-}\nonumber\\&=&{p_{cv}^2\over 6} \sin^2\vartheta,
\end {eqnarray}
where subscripts 1 - 4 correspond to excitonic states (26).
Subscripts $ + $ and $ - $ describe the circular polarization $
{\bf e} _ {{\bf k}} ^ \pm = ({\bf e} _ {x ^\prime} \pm i{\bf e} _
{y ^\prime})/\sqrt {2} $, if the vector $ {\bf k} $ is directed
along $z ^\prime, \vartheta $ is the angle between the
crystallographic axis $z $ and $ {\bf k} $. Summarizing $S _
{\eta\mu} $ on polarizations, we obtain
\begin {eqnarray}
\label {41} &&S _ {1} =S _ {2} = {p _ {cv} ^2\over 3} (1
+\cos^2\vartheta),\nonumber\\ &&S _ {3} =S _ {4} = {p _ {cv}
^2\over 3} \sin^2\vartheta,
\end {eqnarray}
where $S_\eta =\sum_\mu | {\bf p} _ {cv\eta} {\bf e} _ {\mu {\bf
k}} | ^2 $. Substituting (41) in (21), we obtain
\begin {eqnarray}
\label {42}
 \gamma_{r1}&=&\gamma_{r2}={e^2\omega_\eta^3p_{cv}^2\nu\over
6\pi\hbar m_0^2\omega_g^2c^2}\nonumber\\&\times& \int do _ {{\bf
k}} (1 +\cos^2\vartheta) |P ({\bf k} ) | ^2,\nonumber \\
 \gamma_{r3}&=&\gamma_{r4}=\gamma_{r1}/2,~~k=\omega_0\nu/c.
\end {eqnarray}
In a case (36), integrating on angles determining a direction of $
{\bf k}$, we have
\begin {equation}
\label {43} \gamma _ {r1} = \gamma _ {r2} = {8\over 9} {
e^2\omega_\eta^3p _ {cv} ^2\nu\over \hbar m_0^2\omega_g^2c^3} |
{\cal P} (kR) | ^2,
\end {equation}
\begin {equation}
\label {44} \gamma _ {r3} = \gamma _ {r4} = \gamma _ {r1} /2.
\end {equation}
In a case $ k R\ll 1 $, for radiative broadenings of any excitons
$ \Gamma_6\times\Gamma_7 $, we obtain Eq. (43), in which the
factor $ | {\cal P} (k R) | ^2 $ is replaced by $ |P (0) | ^2 $.
At $P (0) =1 $, (43) coincides with Eq. (6) of [9], if in the last
to put $ \varepsilon_1 =\varepsilon_2 =\nu^2. $

\section{Results}
 With the help of the quantum perturbation theory in the
lowest order on the light-electrons interaction, the probability
of the elastic light scattering on low-dimensional size-quantized
semiconductor heterostructures (Eq. (12)) is calculated.  For a
quantum dot, the obtained expression allows to calculate
dimensionless light reflection coefficient, in a case quantum
wires the light scattering section on a length unit. For a quantum
dot the dimension of cross section is $cm^2$.

The differential cross section of resonant light scattering on any
exciton in a quantum dot of any form, potential configuration and
sizes (17) is obtained.

Under condition $ k R\ll 1 $, the polarization and angular
distribution of scattered light do not depend on the quantum dot
form and on the exciton envelope wave function, but depend only on
vectors $ {\bf p} _ {cv\eta} $, which are
 non-diagonal matrix elements of the momentum operator of excitonic
 states. The magnitude of the scattering cross section does not
depend on quantum-dot sizes.

With the help of the quantum perturbation theory, the formula (20)
for the radiative broadening of the exciton $ \eta $ is obtained.
It is applicable to any exciton in any quantum well, quantum wire
and quantum dot at any magnitudes of $ k_\eta R $, where $R $ is
the quantum-well width, quantum-wire diameter or quantum-dot size,
$ k_\eta =\omega_\eta\nu/c, \hbar\omega_\eta $ is the exciton
energy, $ \nu $ is the light refraction factor, which is
considered identical inside
 and outside of the object. For a quantum dot at $ k_\eta R\ll 1 $, we find that the radiative
broadening does not depend on the quantum-dot sizes (21).

The estimation of the magnitude of the total light scattering
cross section in the resonance $ \omega_l = {\tilde \omega} _0 $
shows that at $ \gamma\ll\gamma_r $, the section is about $
(\lambda_l/2\pi) ^2x $, where $x $ is some number.

The example of the exciton $ \Gamma_6\times\Gamma_7 $ in cubic
crystals $T_d $ is considered.

\end {document}